\documentclass[aps,pre,reprint,superscriptaddress]{revtex4-2}

\usepackage{amsmath}
\usepackage{amssymb}
\usepackage{hyperref}

\usepackage{graphicx}

\begin{document}

%\preprint{}

\title{Suppression of Quasiperiodicity in Circle Maps with Quenched Disorder}


\author{David Müller-Bender}
\email[]{david.mueller-bender@mailbox.org}
\author{Johann Luca Kastner}
\email[]{luca.kastner@gmx.de}
\affiliation{Institute of Physics, Chemnitz University of Technology, 09107 Chemnitz, Germany}
\author{Günter Radons}
\email[]{radons@physik.tu-chemnitz.de}
\affiliation{Institute of Physics, Chemnitz University of Technology, 09107 Chemnitz, Germany}
\affiliation{Institute of Mechatronics, 09126 Chemnitz, Germany}
%\thanks{}
%\affiliation{}


\date{\today}

\begin{abstract}
We show that introducing quenched disorder into a circle map leads to the suppression of quasiperiodic behavior in the limit of large system sizes. Specifically, for most parameters the fraction of disorder realizations showing quasiperiodicity decreases with the system size and eventually vanishes in the limit of infinite size, where almost all realizations show mode-locking. Consequently, in this limit, and in strong contrast to standard circle maps, almost the whole parameter space corresponding to invertible dynamics consists of Arnold tongues.
\end{abstract}

%\keywords{}

\maketitle

% body of paper here
Circle maps can be considered as Poincaré maps of periodically forced nonlinear oscillators \cite{pikovsky_synchronization:_2001}.
As simple models for this general concept they appear in various fields such as biology \cite{glass_simple_1979,neiman_synchronization_1999}, physiology \cite{glass_bifurcation_1983,arnold_cardiac_1991,glass_cardiac_1991}, semiconductor physics \cite{martin_circle_1986,gwinn_frequency_1986,*gwinn_frequency_1986_erratum1,*gwinn_frequency_1986_erratum2}, fluid dynamics \cite{stavans_fixed_1985,*stavans_fixed_1985_erratum,jensen_global_1985,olinger_nonlinear_1988}, thermoacoustics \cite{yazaki_nonlinear_1990}, electrodynamics \cite{hoffnagle_frequency-locked_1993}, optics \cite{de_la_llave_theory_1999}, quantum mechanics \cite{petrov_dynamical_2005}, and in the theory of deterministic ratchets \cite{salgado-garcia_deterministic_2006}.
They play a crucial role in the theory of time-delay systems with time-varying delay \cite{otto_universal_2017,muller_dynamical_2017}, where they describe the dynamics of the feedback of the system.
Depending on the circle map dynamics, the latter systems show fundamentally different types of chaos \cite{muller_laminar_2018,muller-bender_resonant_2019,hart_laminar_2019,muller-bender_laminar_2020,jungling_laminar_2020,kulminskii_laminar_2020} and their solutions have different analyticity properties \cite{mallet-paret_analyticity_2014}.
The rich variety of dynamics of circle maps was extensively studied in the literature.
In the pioneering work of Poincaré \cite{poincare_sur_1885}, Denjoy \cite{denjoy_sur_1932}, and Arnold \cite{arnold_small_1961, *arnold_small_1961_erratum}, fundamental properties of monotonically increasing circle maps, especially homeomorphisms and diffeomorphisms of the circle, such as the appearance of quasiperiodicity and mode-locking dynamics as well as the existence of a topological conjugacy to the pure rotation, were elaborated.
Beyond monotonically increasing circle maps, further studies investigated the transition between regular and chaotic dynamics \cite{glass_fine_1982,jensen_complete_1983,cvitanovic_renormalization_1985,mackay_transition_1986,arneodo_crossover_1987,ding_scaling_1987,artuso_phase_1989} as well as deterministic diffusion \cite{geisel_onset_1982,geisel_accelerated_1985,groeneveld_negative_2002,korabel_fractal_2002,artuso_anomalous_2003}.
Also systems of mutually coupled circle maps were considered, where, for instance, phase synchronization \cite{osipov_phase_2002} and spatio-temporal intermittency \cite{stassinopoulos_measuring_1990} can be observed.
While the response of circle maps to external perturbations, such as coupling with a chaotic map \cite{pikovsky_attractor-repeller_1997} as well as quasiperiodic \cite{cartwright_universality_1999} and stochastic forcing \cite{neiman_synchronization_1999,kleptsyn_contraction_2004,zmarrou_dynamics_2008,rodrigues_family_2015,malicet_random_2017,malicet_lyapunov_2021,marangio_arnold_2020}, was extensively studied, the influence of quenched disorder on the dynamics of circle maps by now is hardly understood.
Disordered circle maps arise naturally when one studies delay systems with randomly varying delay, a very common situation \cite{verriest_stability_2009,krapivsky_stochastic_2011,gomez_stability_2016,qin_stability_2017,liu_stability_2019}.
Their mode-locking behavior is of fundamental importance for the classification of these systems \cite{otto_universal_2017,muller_dynamical_2017}.
The prototypical circle map of Arnold arises also in the large dissipation limit of the dissipative standard map \cite{zaslavsky_simplest_1978,schmidt_dissipative_1985}, which in turn is a Poincaré map of a damped particle, kicked, or driven periodically, with a spatially periodic potential.
Replacing the latter by a spatially random potential leads in the same limit to the systems studied in this paper.
While spatial randomness is a fundamental concept in solid state physics \cite{lifshits_introduction_1988,luck_lifshitz_1988,luck_cantor_1989}, it is largely unexplored for dynamical systems although it is known that it can change the dynamics drastically (for instance, see \cite{radons_suppression_1996,stiller_dynamics_1998,acebron_kuramoto_2005,zech_dynamics_2020,zech_dynamics_2021}).
Results for the special case of expanding circle maps and general expanding dynamical systems, which is not considered here, can be found in \cite{simula_deterministic_2009,stenlund_coupling_2013} and \cite{aimino_annealed_2015}, respectively.
In \cite{le_spatially_2015}, quenched disorder is introduced by a adding a random function to a monotonically increasing circle map, which destroys monotonicity leading to a destabilizing effect of the disorder.
\begin{figure}
	\includegraphics[width=0.48\textwidth]{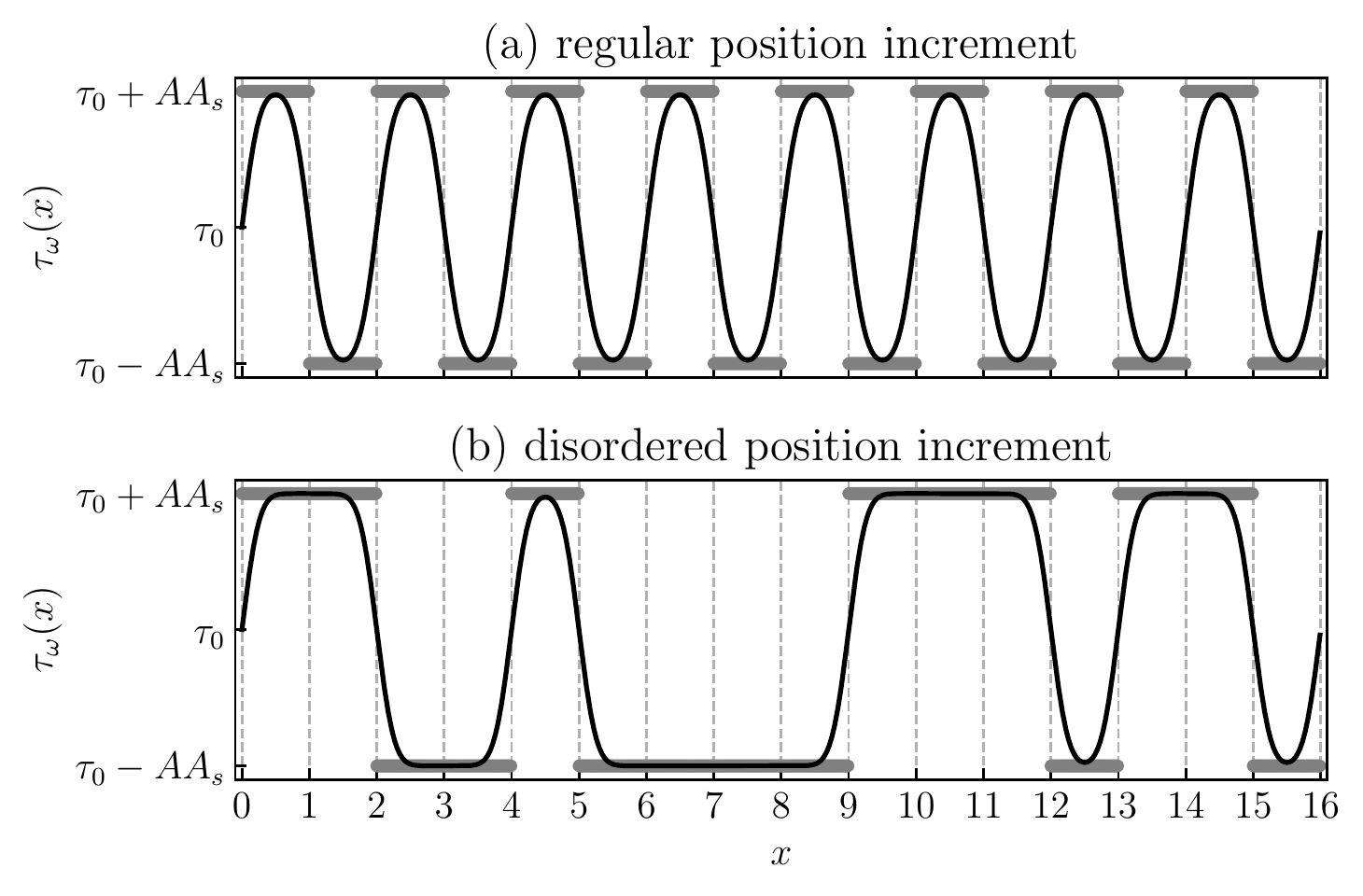}
	\caption{Variation of the position increment $\tau _{\omega}(x)$ (black line) per iteration of the circle map, Eq.~\eqref{eq:rotangle}, as function of the current position $x$ on the circumference of a circle (perimeter $L=16$), (a) for a regular and (b) for a disordered circle map.
		In both cases, $\tau _{\omega}(x)$ is obtained by a Gaussian smoothing of a dichotomic signal (grey line), which alternates periodically in (a) and irregularly in (b).
		For large $L$ this difference entails drastically different mode-locking behavior as shown in Fig.~\ref{fig:arnold}.
	}
	\label{fig:rotangle}
\end{figure}
In this Letter, we investigate disordered monotonically increasing circle maps, where we focus on the finite size scaling of the disorder averages of the drift velocity and the Lyapunov exponent.
We demonstrate that these quantities are well defined also in the limit of an infinite system size, where, almost surely, a unique limit value is reached.
Finally, it is shown that the fraction of disorder realizations leading to quasiperiodic dynamics decays with increasing system size and vanishes in the limit of an infinite size, if all position increments exceed the spatial correlation length of the disorder.

We consider one-dimensional iterated maps of the form
\begin{equation}
	\label{eq:extended_map}
	x_{n+1}=R(x_{n})=x_{n}+\tau (x_{n}),
\end{equation}
where $\tau (x)>0$, a position increment at position $x\in\mathbb{R}$, is $L$-periodic, i.e., $\tau(x+L)=\tau(x)$.
Such maps are lifts of circle maps \cite{katok_introduction_1997}.
The standard example is the Arnold family \cite{arnold_small_1961} with $L=1$ and $\tau(x)=\tau _{0}+AA_{s}\sin (2\pi x)$, where the factor $A_{s}=1/(2\pi )$ guarantees that $R(x)$ is monotonically increasing if $0\leq A\leq 1$.
In this Letter, we will consider spatially disordered versions of Eq.~\eqref{eq:extended_map} with $\tau(x)$ replaced by $\tau_\omega(x)$, where $\omega$ specifies a disorder realization.
Our strategy consists of introducing disorder in an interval of size $L$, which is then repeated periodically, and eventually the size $L$ of such a unit cell is sent to infinity.
We can then, for all finite sizes $L$, rely on the known properties of circle maps, while we approach the infinite size limit.
Exemplary for generic smooth random functions, we obtain $\tau_{\omega}(x)$ as a smoothed, scaled and shifted version of a dichotomic signal $\chi_{\omega }(x)$, which is piecewise constant with values $S_{i}=\pm 1$ for $x\in [i-1,i)$ with $i$ integer.
A disorder realization in a size-$L$ unit cell is determined by a symbol chain $\omega=S_{1}S_{2}\ldots S_{L}$, given by $L$ random symbols $S_{i}$, which are independent and equally distributed.
The random function, which replaces the sine in the Arnold family, is then obtained by a convolution $G_{\varsigma }\ast \chi_{\omega}(x)$ with a Gaussian $G_{\varsigma}(x)$ of variance $\varsigma^2=0.04$.
Scaling by $AA_{s}$ and shifting by an offset $\tau_0$ gives our disordered position increment
\begin{equation}
	\label{eq:rotangle}
	\tau_{\omega}(x)=\tau_{0}+AA_{s}G_{\varsigma}\ast \chi_{\omega}(x),
\end{equation}
where $A_{s}=\sqrt{2\pi\varsigma ^{2}}/2$ guarantees monotonicity of $R_\omega(x)=x+\tau_\omega(x)$ for amplitudes $A$ with $0\leq A\leq 1$.
Applying periodic boundary conditions for $\tau_\omega(x)$ results in our circle map with quenched disorder $\omega$,
\begin{equation}
	\label{eq:circle_map}
	x_{n+1}= R_{\omega}(x_{n})\; \text{mod}\; L=[x_{n}+\tau_{\omega}(x_{n})]\; \text{mod}\; L,
\end{equation}
which is treated in the rest of this Letter.
Fig.~\ref{fig:rotangle} shows two examples of $\tau_{\omega }(x)$ of Eq.~\eqref{eq:rotangle}, one for a regular configuration $\omega =(+-+-\ldots +-)$ and one for the disordered case $\omega =(++--+----+++-++-)$ with $L=16$, together with the shifted and scaled version of $\chi _{\omega }(x)$.
Our main results below are valid for parameters $A\in [0,1]$ and arbitrary $\tau_0$, which are kept fixed, i.e., independent of the system size $L$, while $L$ is varied.
This especially implies that, in the scaling limit $L\to\infty$, we can assume that $L\gg \tau_0$
\footnote{A different scaling limit $L\rightarrow \infty $ is obtained, if $\tau_0$ increases simultaneously with $L$, e.g. for $\tau_0(L)\sim L$.
We have strong numerical evidence that also in this limit almost all disorder realizations for almost all parameters lead to mode-locking.}.

\begin{figure*}
	\includegraphics[width=0.99\textwidth]{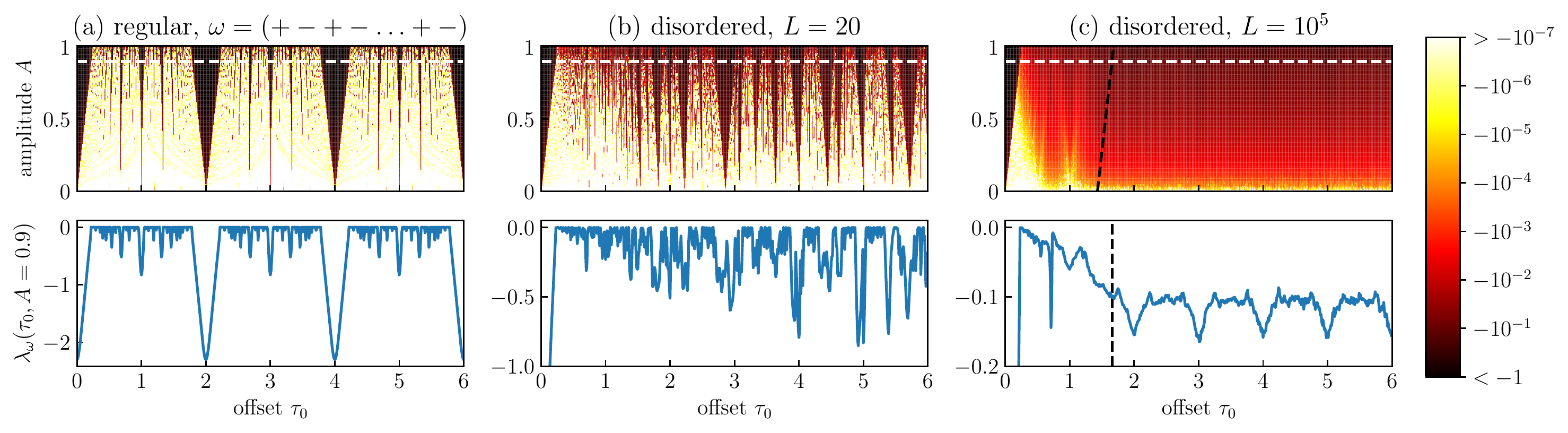}
	\caption{Heat maps $\lambda _{\omega }(\tau _{0},A)$ of the Lyapunov exponent (top panels)
		in dependence of the offset $\tau _{0}$ and the amplitude $A$ (see Eq.~\eqref{eq:rotangle}) are shown in (a) for the periodic configuration of Fig.~\ref{fig:rotangle}a ($L$ even, otherwise arbitrary), and for an individual disorder configuration as in Fig.~\ref{fig:rotangle}b, for $L=20$ (b), and $L=10^{5}$ (c).
		While (b) looks like a	disordered version of (a), we find in (c) an apparent separation into two regimes: to the right of the dashed black line, where all position increments $\tau_{\omega}(x)$ are larger than the correlation length of $\tau_{\omega}(x)$ \cite{Note3}, mode-locking ($\lambda_{\omega }(\tau _{0},A)<0$) is found for the overwhelming majority of parameters.
		We will argue that	this holds for \emph{almost all} parameters for $L\rightarrow \infty $. In the bottom panels, this picture is confirmed for $\lambda_{\omega}(\tau _{0},A=0.9)$.
	}
	\label{fig:arnold}
\end{figure*}

It is known that monotone circle maps exhibit two types of dynamics \cite{katok_introduction_1997}, which can be classified by the rotation number \cite{katok_introduction_1997}
\begin{equation}
	\label{eq:rotationnumber}
	\rho_\omega = \lim_{N\to\infty} \frac{x_N-x_0}{N\,L} = \lim_{N\to\infty} \frac{1}{N\,L} \sum_{n=0}^{N-1} \tau_\omega(x_n)
\end{equation}
and by the Lyapunov exponent \cite{ott_chaos_2002}
\begin{equation}
	\label{eq:lyapunovexponent}
	\lambda_\omega=\lim_{N\to\infty}\frac{1}{N} \sum_{n=0}^{N-1} \ln(1+\tau'_\omega(x_n)).
\end{equation}
So-called mode-locking dynamics is characterized by attracting periodic orbits.
In this case, the rotation number $\rho_\omega$ is rational and the Lyapunov exponent is negative, $\lambda_\omega < 0$.
In contrast, quasiperiodic dynamics is characterized by marginally stable quasiperiodic motion with $\lambda_\omega=0$ and the rotation number $\rho_\omega$ is irrational \footnote{
While a vanishing Lyapunov exponent not necessarily implies quasiperiodic dynamics, a negative Lyapunov exponent implies mode-locking dynamics given that the circle map is noncritical, $0 < A < 1$, and is continuously differentiable \cite{de_faria_real_2016}.
As shown in the latter reference, this also holds for critical circle maps, $A=1$, if they are 3-times continuously differentiable.
Since our map defined by Eqs.~(\ref{eq:rotangle},\ref{eq:circle_map}) is even smooth, a negative Lyapunov exponent implies mode-locking for $0 < A \leq 1$.
}.
To illustrate in parameter space the effect of disorder on these types of dynamics, we computed the heat maps of the Lyapunov exponent $\lambda_\omega$, also called Lyapunov graphs \cite{de_figueiredo_lyapunov_1998}, shown in Fig.~\ref{fig:arnold}.
We considered a regular map, where $\omega$ is periodic, and two realizations of disordered maps, comparing the effect of small ($L=20$) and large ($L=10^5$) system sizes.
For the regular map, periodic structures known from the Arnold family \cite{arnold_small_1961,arnold_small_1961_erratum} appear in the parameter space, where the connected parameter sets related to mode-locking dynamics with $\lambda_\omega<0$ are called Arnold tongues (Fig.~\ref{fig:arnold}a).
The remaining sets, which lead to quasiperiodic dynamics with $\lambda_\omega=0$, are fat fractals (for $A<1$), which are of nonzero measure \cite{ott_chaos_2002}.
For the disordered map with system size $L=20$, the situation is nearly the same with the difference that the period of the structures in parameter space is equal to the system size $L$ and thus larger than the period for the regular map shown in Fig.~\ref{fig:rotangle}a.
This leads to the disordered appearance of the structures in the considered subset of the parameter space (Fig.~\ref{fig:arnold}b).
For large system size, $L=10^5$, the structure of the parameter space changes drastically, as it separates into basically two characteristic regions (Fig.~\ref{fig:arnold}c), which are roughly separated by $\tau_0(A)=l+A\,A_s$ (dashed black line), where the minimal position increment $\tau_{\text{min}} = \min_{\omega,x} \tau_\omega(x) =\tau_0 - A\,A_s$ equals the spatial correlation length of the disorder $l\approx 1.43268$
\footnote{We define the correlation length $l>0$ by $C(\Delta x=l)/C(\Delta x=0)=0.01$, where $C(\Delta x) = \delta_{x=\Delta x}^2 [ (\varsigma/\sqrt{\pi}) \exp(- x^2/(4\varsigma^2)) + (x / 2) \, \mathrm{erf}(x / (2\varsigma)) ] $, with $\delta_{x=x'}^2[f(x)] = f(x'+1) - 2 f(x') + f(x'-1)$, is proportional to the spatially averaged covariance function of $\tau_\omega(x)$ for $L=\infty$.}.
In the region defined by $\tau_{\text{min}}<l$, fractal structures similar to the ones observed for the regular and the disordered map with small $L$ appear, indicating that there both types of dynamics have nonzero measure.
In stark contrast, for $\tau_{\text{min}}>l$, where the position increment $\tau_\omega(x)$ exceeds the correlation length of the disorder, the fractal structures known from regular circle maps are replaced by regular periodic structures and it seems that almost all parameters in this region lead to mode-locking dynamics with $\lambda_\omega<0$.
This seems to contradict the known fact that the measure of the parameter sets for quasiperiodic dynamics is nonzero.
In fact, for large but finite $L$, the measure is nonzero but extremely small and eventually vanishes in the limit $L\to\infty$ as we demonstrate below. 
 
\begin{figure}
	\includegraphics[width=0.48\textwidth]{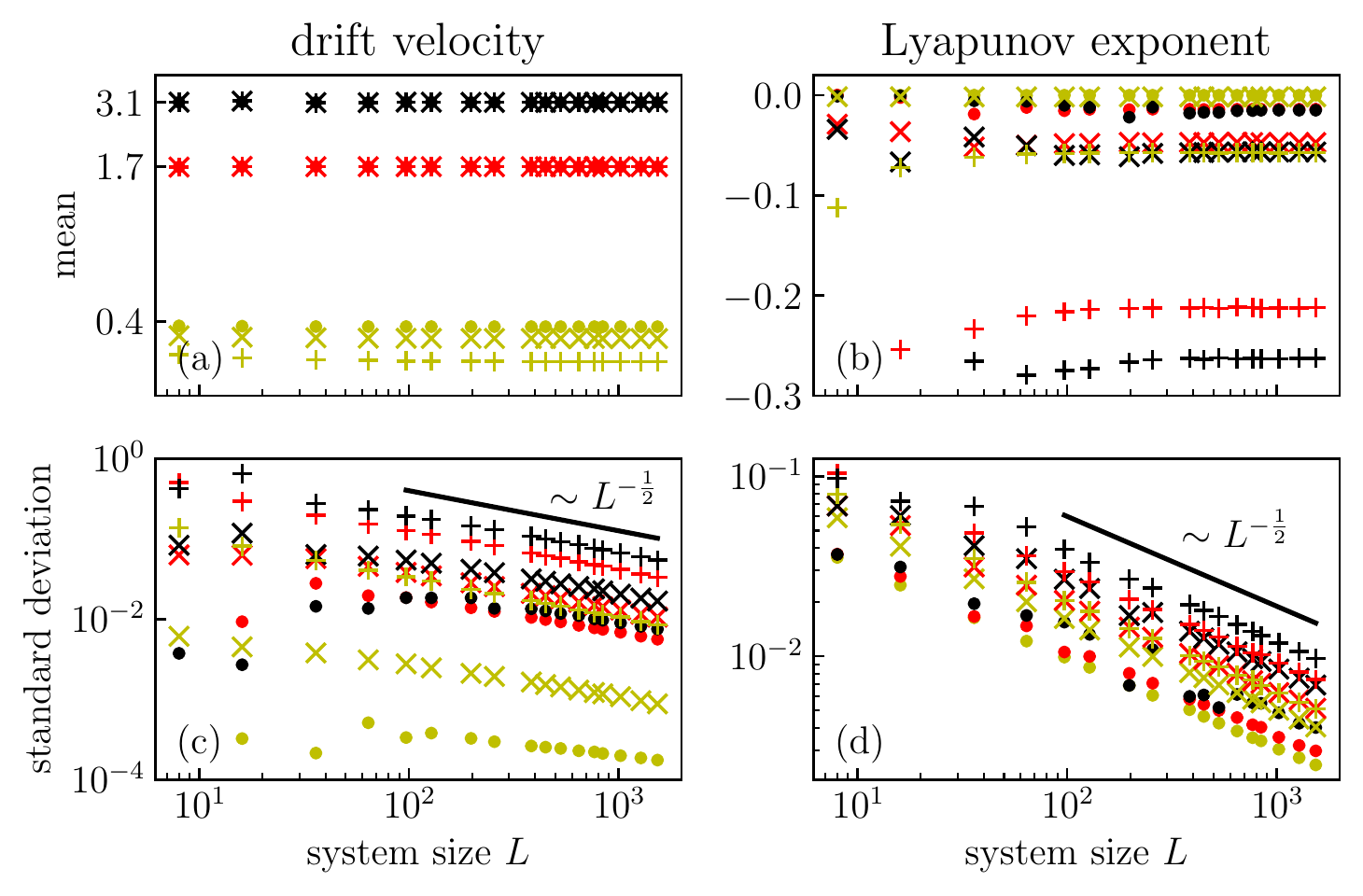}
	\caption{Disorder averaged mean (top panels) and standard deviation (bottom panels) of the drift velocity $v_\omega$ (left panels) and the Lyapunov exponent $\lambda_\omega$ (right panels) as function of the system size $L$ for various parameters $\tau_0$ and $A$.
		The offset $\tau_0$ was set to $0.4$ (yellow), $1.7$ (red), and $3.1$ (black).
		The amplitude $A$ was set to $0.4$ (dots), $0.7$ (crosses), and $1.0$ (pluses).
		In the limit of large $L$, the standard deviation of both quantities decreases asymptotically with $L^{-1/2}$, whereas the mean approaches a limit value.
		Therefore, for $L\to\infty$, Lyapunov exponent and drift velocity converge almost surely to their disorder average, and thus for $L=\infty$ almost all realizations of the map defined by Eqs.~(\ref{eq:rotangle},\ref{eq:circle_map}) have the same drift velocity and Lyapunov exponent for given $\tau_0$ and $A$.
	}
	\label{fig:driftlyaconv}
\end{figure}

To show that the observations we made above for single disorder realizations are representative for a generic ensemble of disorder realizations, we analyze the limiting behavior for $L\to\infty$ of the Lyapunov exponent given by Eq.~\eqref{eq:lyapunovexponent} and the drift velocity $v_\omega$, which is defined by
\begin{equation}
	\label{eq:drift}
	v_\omega = L\, \rho_\omega,
\end{equation}
where $\rho_\omega$ is the rotation number given by Eq.~\eqref{eq:rotationnumber}.
In Fig.~\ref{fig:driftlyaconv} the disorder averaged mean and standard deviation of the drift velocity $v_\omega$ and the Lyapunov exponent $\lambda_\omega$ are plotted as function of the system size $L$, where the average was computed from $10^4$ disorder realizations.
For $L\to\infty$, the means converge to limit values, while the standard deviations asymptotically vanish as $L^{-1/2}$.
For fixed $\tau_0$ and $A$, this means that there is a unique drift velocity $v_\omega \to v_\infty$ and a unique Lyapunov exponent $\lambda_\omega \to \lambda_\infty$ in the limit $L\to\infty$, which are observed for almost all realizations.
Moreover, it follows that the values obtained for a large system are well approximated by the ensemble average of (smaller but also large) subsystems, i.e., $v_\omega$ and $\lambda_\omega$ are self-averaging.

\begin{figure}
	\includegraphics[width=0.48\textwidth]{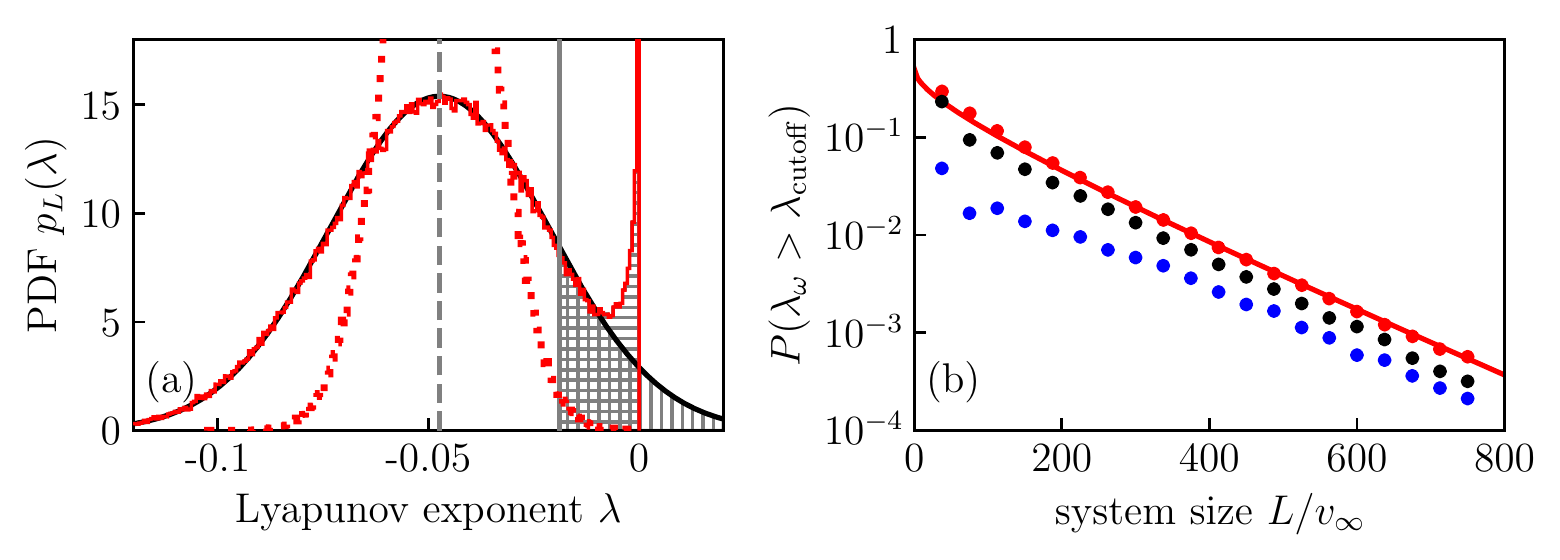}
	\caption{The fraction of disorder realizations that lead to quasiperiodic dynamics, where we have $\lambda_\omega=0$, vanishes in the limit $L\to\infty$ if $\tau_{\text{min}}>l$:
		(a) Distribution of the Lyapunov exponent $p_L(\lambda)$ with respect to the disorder (red line) and its Gaussian approximation (black line, see text) for $A=0.7$, $\tau_0=1.7$, and $L = 255 \approx 150\, v_\infty$.
		For a fixed $\lambda_{\text{cutoff}}$ (solid gray vertical line), $P(\lambda_\omega > \lambda_{\text{cutoff}})$ and its approximation by Eq.~\eqref{eq:fracquasiperio} are given by the hatched areas below the red and black line, respectively.
		(b) Scaling behavior with respect to the system size $L$ of the fraction of disorder realizations that lead to quasiperiodic dynamics.
		Numerical estimates of $P(\lambda_\omega > \lambda_{\text{cutoff}})$ with $\lambda_{\text{cutoff}}=\lambda_\infty/5$ (discrete marks) are compared to the analytical estimate $\hat{P}(\lambda_{\omega }>\lambda_{\text{cutoff}})$ given by Eq.~\eqref{eq:fracquasiperio} (solid line).
		We have set the amplitude $A=0.7$, where from top to bottom the offset $\tau_0$ equals $1.7$ (red), $3.1$ (black), and $9.1$ (blue).
	}
	\label{fig:fracquasiperio}
\end{figure}

Specifically, this implies that the underlying density of Lyapunov exponents $p_{L}(\lambda)=\overline{\delta (\lambda - \lambda_{\omega})}$, where the overline denotes the disorder average, approaches for $L\to\infty$ a $\delta$-distribution, $p_{\infty}(\lambda)=\delta(\lambda -\lambda_{\infty})$.
The question then is, how and under which conditions this limit is approached. The first point is answered numerically in Fig.~\ref{fig:fracquasiperio}a for a case with negative Lyapunov exponent $\lambda_{\infty}$ for two system sizes $L=255$ and $L=1275$.
By comparing with a Gaussian $G_{\mu,\sigma(L)}(\lambda)$, whose mean $\mu$ and variance $\sigma^{2}(L)$ is derived below, we see that for increasing $L$ this fits the numerical results well near the center $\lambda_{\infty}$.
Of course, there have to be deviations near $\lambda=0$, because the true distribution has to vanish identically for $\lambda>0$.
But for cases with $\lambda_{\infty}<0$ also these deviations vanish with increasing system size $L$, because the total probability of finding a Lyapunov exponent $\lambda_{\omega}$ near $\lambda=0$ vanishes as demonstrated numerically in Fig.~\ref{fig:fracquasiperio}b via the integrated density $P(\lambda_{\omega}>\lambda_{\text{cutoff}})=\int_{\lambda_{\text{cutoff}}}^{\infty }p_{L}(\lambda)\; d\lambda$, where $\lambda_{\text{cutoff}}$ is an arbitrary cut-off value larger than $\lambda_{\infty }$, i.e. $\lambda_{\infty}<\lambda_{\text{cutoff}}<0$.
For all of the chosen model parameters $P(\lambda_\omega > \lambda_{\text{cutoff}})$ decays exponentially with system size $L$, which supports our conjecture on the suppression of quasiperiodicity in the limit $L\to\infty$ for $\tau_{\text{min}}>l$.
In this figure we also display the analytical estimate $\hat{P}(\lambda_{\omega }>\lambda_{\text{cutoff}})=\int_{\lambda_{\text{cutoff}}}^{\infty} G_{\mu ,\sigma (L)}(\lambda )\; d\lambda$, Eq.~\eqref{eq:fracquasiperio}, for the probability $P(\lambda_{\omega}>\lambda_{\text{cutoff}})$, based on the Gaussian approximation $G_{\mu,\sigma(L)}(\lambda)$ and see that this gives a good bound.
All averages in this paragraph were obtained from $320000$ disorder realizations. 

We finally discuss why $\lambda_{\infty}$ is strictly negative in the limit $L\to\infty$ for $\tau_{\text{min}}>l$ and validate our arguments by deriving from them the correct large-$L$ asymptotics of $P(\lambda_\omega > \lambda_{\text{cutoff}})$.
For that, we consider the finite-time Lyapunov exponent $\lambda_{\omega,N} = \frac{1}{N} \sum_{n=0}^{N-1} \log(1+\tau'_\omega(x_n))$ of an orbit that passed through the system once.
In detail, we consider $N=q_\omega$, where $q_\omega$ is the largest natural number that fulfills $x_{q_\omega}-x_0<L$ and argue that, in the limit $L\to\infty$, $\lambda_{\omega,N=q_\omega}$ is strictly negative and that we have $\lambda_{\omega,N=q_\omega} \to \lambda_{\omega}=\lambda_{\omega,N=\infty}$, which implies $\lambda_\omega < 0$.
Provided that $\tau_{\text{min}}>l$, the increments $\tau_\omega(x_{n})$ and $\tau_\omega(x_{n+1})$ are nearly independent, where independence strictly holds if the Gaussian $G_\varsigma(x)$ in Eq.~\eqref{eq:rotangle} is replaced by a distribution with finite support $[-l,l]$, i.e., the tails of the Gaussian are neglected.
For $0\leq n\leq q_\omega$, our disordered circle map then is stochastically equivalent to the circle map $x_{n+1}=r_{\omega_n}(x_n)=[x_n+\tau_{\omega_n}(x_n)] \mod L$ with annealed disorder, known as ``random dynamical system on the circle'' \cite{kleptsyn_contraction_2004,zmarrou_dynamics_2008,rodrigues_family_2015,malicet_random_2017,malicet_lyapunov_2021}, where the map randomly changes with each iteration, since the state in our system with quenched disorder ``sees'' with each iteration an independent disorder realization. 
For random dynamical systems on the circle it is known that the Lyapunov exponent is strictly negative if the maps $r_{\omega_n}$, almost surely, do not preserve the same measure \cite{malicet_random_2017,malicet_lyapunov_2021}, a very general assumption.
So we can assume that, in the limit $L\to\infty$, $\lambda_{\omega,q_\omega}$ converges to a strictly negative value for almost all $\omega$ and arbitrary initial values $x_0$.
It follows that almost every orbit of length $q_\omega$ is attracted by an attractive orbit.
Assuming that the latter is a period $q_\omega$ orbit $\{x^*_0,x^*_1,\dots,x^*_{q_{\omega}-1}\}$, in the limit $L\to\infty$, $\lambda_{\omega,q_\omega}$ converges to the Lyapunov exponent $\lambda_{\omega} = \frac{1}{q_\omega} \sum_{n=0}^{q_\omega-1} \log(1+\tau'_\omega(x^*_n))$ of this periodic orbit since $q_\omega \approx L/v_\infty$ grows unboundedly with $L$ so that deviations for arbitrary initial values $x_0\neq x^*_n$ caused by the transients vanish in the limit $L\to\infty$.
Due to the independence of the increments $\tau_\omega(x_n)$, $\lambda_{\omega}$ then is an arithmetic mean of $q_\omega$ independent random variables, and thus, according to the central limit theorem, $p_L(\lambda)$ can be approximated by a Gaussian distribution $G_{\mu,\sigma(L)}(\lambda)$ with mean $\mu=\lim_{L \to\infty} \overline{\lambda_{\omega,q_\omega}}$.
The variance $\sigma^2(L)$ is inversely proportional to the number $q_\omega \approx L/v_\infty$ of random variables.
In detail, we have $\sigma^2(L) = (v_\infty/L)\,\sigma_0^2$, with $\sigma_0^2=\lim_{L \to\infty} (L/v_\infty)\,\overline{(\lambda_{\omega,q_\omega}-\lambda_\infty)^2}$, which confirms the $L$-dependence observed in Fig.~\ref{fig:driftlyaconv}d.
In practice, $\mu$ and $\sigma_0^2$ can be numerically approximated from disorder averages of the finite-time Lyapunov exponent $\lambda_{\omega,N}$, with large $N$, of the system with $L=\infty$, by dropping the limit in their equations and substituting $q_\omega$ and $L/v_\infty$ with $N$.
Integrating $G_{\mu,\sigma(L)}(\lambda)$ over all $\lambda \in [\lambda_{\text{cutoff}},\infty)$ gives an analytical estimate for $P(\lambda_\omega > \lambda_{\text{cutoff}})$,
\begin{eqnarray}
	\hat{P}(\lambda_\omega > \lambda_{\text{cutoff}})
	&=& \frac{1}{2} (1-\mathrm{erf}(c\,\sqrt{L/v_\infty})) \label{eq:fracquasiperio}\\
	&\sim& (L/v_\infty)^{-1/2}\,e^{-c^2\,L/v_\infty} \label{eq:fracquasiperioasym},
\end{eqnarray}
where we have $c=(\lambda_{\text{cutoff}}-\lambda_\infty)/\sqrt{2\,\sigma_0^2}$, and the asymptotic expansion of the error function (cf. \cite{abramowitz_handbook_1972}) in Eq.~\eqref{eq:fracquasiperioasym} holds for $\lambda_{\text{cutoff}}>\lambda_\infty$.
In Fig.~\ref{fig:fracquasiperio}b, Eq.~\eqref{eq:fracquasiperio} is compared to numerical estimates of $P(\lambda_\omega > \lambda_{\text{cutoff}})$ for fixed $A=0.7$ and different $\tau_0$ with $\tau_{\text{min}}>l$.
Eq.~\eqref{eq:fracquasiperio} is plotted only for $\tau_0=1.7$ since the values of $c$ are nearly identical for the considered $\tau_0$.
While Eq.~\eqref{eq:fracquasiperio} reproduces the asymptotic decay of $P(\lambda_\omega > \lambda_{\text{cutoff}})$ for all considered $\tau_0$ in the limit of large $L$, the numerical estimates of $P(\lambda_\omega > \lambda_{\text{cutoff}})$ differ from Eq.~\eqref{eq:fracquasiperio} by a factor $\alpha \leq 1$, which we found to decrease for increasing $\tau_0$.

In this Letter, we have analyzed the dynamics of disordered circle maps.
As a representative example, the position increment was defined by a disordered smoothed dichotomic signal.
Focusing on monotonically increasing circle maps with quenched disorder, we have shown that the fraction of disorder realizations that lead to marginally stable quasiperiodic dynamics decreases with increasing system size $L$ and eventually vanishes in the limit $L\to\infty$ for almost all parameters provided that the position increment exceeds the spatial correlation length of the disorder.
It will be interesting to see, whether in this limit the remaining set of parameters with quasiperiodic behavior has fractal dimensions analogous to the Arnold family for $A \to 1$ \cite{jensen_complete_1983,cvitanovic_renormalization_1985,arneodo_crossover_1987,artuso_phase_1989}.
We further have demonstrated that the drift velocity and the Lyapunov exponent almost surely approach well defined limiting values in the limit $L\to\infty$.
Our results are expected to be valid more generally for monotone circle maps with short-ranged correlations of the disorder.

\begin{acknowledgments}
	D.MB. and G.R. gratefully acknowledge funding by the Deutsche Forschungsgemeinschaft (DFG, German Research Foundation) - 456546951.
\end{acknowledgments}

\bibliography{disorderedcirclemaps-prl}

\end{document}